\documentclass[%
 reprint,
superscriptaddress,
nofootinbib,
 amsmath,amssymb,
 aps,
]{revtex4-1}

\usepackage{graphicx}
\graphicspath{ {./Figures/} }
\usepackage{dcolumn}
\usepackage{bm}
\usepackage{tablefootnote} 
\usepackage{threeparttable} 
\usepackage{physics}
\usepackage{xcolor}
\usepackage{soul}
\usepackage{hyperref}
\usepackage{mathtools}
\usepackage{bm}

\newcommand{\einf}{\epsilon_{\infty}}

\newcommand{\expqd}{e^{qd}}

\begin{document}


\title{Generalized deformation potential and machine-learning approaches for electron-phonon coupling and thermoelectric transport in semiconductors} 

\author{Ransell D'Souza}
\email{rdsouza@sissa.it}
\affiliation{
Scuola Internazionale Superiore di Studi Avanzati (SISSA), I-34136 Trieste, Italy}

\author{Ivana Savi\'c}
\email{ivana.savic@kcl.ac.uk}
\affiliation{%
Department of Physics, King's College London, The Strand, London WC2R 2LS, United Kingdom
}%

\date{\today}

\begin{abstract}

The ability to compute electron-phonon coupling from first principles, using density functional perturbation theory and interpolation techniques, has enabled predictive calculations of electronic transport coefficients in crystalline materials. However, these methods are still computationally expensive. 
Here we present two inexpensive methods to obtain thermoelectric transport properties of semiconductors using a small number of electron-phonon matrix elements calculated from first principles. The first method combines models for coupling of electrons with different phonon modes whose parameters are obtained from $\sim 10$ matrix elements per electronic band and phonon mode calculated from first principles. Within this method, we formulate the acoustic deformation potential model for arbitrary crystal symmetries and band extrema locations. The second method uses machine learning to interpolate $\sim 100$ electron-phonon matrix elements per electronic band and phonon mode on dense reciprocal space grids in the parts of the Brillouin zone relevant for transport. We apply both methods to two-dimensional MoS$_2$ and show very good agreement with the 
state-of-the-art method. The calculated thermoelectric properties also agree well with experiments. We find that the machine-learning method is more accurate and straightforward to implement compared to the model approach.
\end{abstract}

\maketitle

\section{Introduction}\label{sec:int}

There has been significant recent progress in the development of methods and  codes to calculate electron-phonon interactions and intrinsic electronic transport properties of bulk and two-dimensional (2D) materials from first principles~\cite{Claes2025,  Ponce2023_PRB, Ponce2021, PhysRevB.101.184302, Lee2023, Zhou2021, Cepellotti2022,Marini2024, ABINIT2025}. These methods combine the Boltzmann transport equation (BTE) with approaches based on density functional theory (DFT) to compute the electronic and phonon band structure and the electron-phonon (el-ph) matrix elements on a reciprocal space grid. In the case of semiconductors, these grids need to be very dense  as the charge carriers contributing to transport occupy a narrow energy range near the Fermi level and, consequently, a small portion of the first Brillouin zone. Calculations on dense grids \textcolor{black}{(of the order of $100$ points per direction in reciprocal space, both for electrons and phonons)} are particularly costly for el-ph matrix elements. To overcome this problem, they are usually computed on a small  grid using density functional perturbation theory (DFPT)~\cite{gonze97,Baroni2001} \textcolor{black}{(of the order of $10$ points per direction in reciprocal space, both for electrons and phonons)} and then interpolated to a much denser grid using the Wannier~\cite{PhysRevB.76.165108} or Fourier~\cite{PhysRevB.102.094308} interpolation. Nevertheless, this approach is still computationally intensive, particularly when used for screening of materials with desired transport properties or predicting transport properties of materials with large unit cells (e.g.~heterostructures).

One approach to reduce the cost of {\it ab-initio} el-ph calculations in semiconductors is to use models for el-ph coupling based on the type of phonons participating in the interaction 
(e.g.~the deformation potential models for acoustic, non-polar optical and zone-edge phonons or the Fr\"ohlich model for polar optical phonons)~\cite{ridley99}. The parameters of those models can be calculated using \textcolor{black}{a small set of} el-ph matrix elements \textcolor{black}{(of the order of 10 matrix elements per electronic band and phonon mode)} obtained from the DFPT approach or its alternatives (e.g.~the DFT-based frozen phonon approach)~\cite{PhysRevB.78.035202, Murphy2018, Cao2018, DSouza2020, Ganose2021, ELECTRA2023}. However, an accurate formulation of such models can be challenging, which is why they are often oversimplified. For example, the acoustic deformation model is frequently used by accounting for only one deformation potential parameter~\cite{Pei2012,Wang2012,Witting2019}, although there can be up to six independent parameters describing the deformation potential tensor depending on  crystal symmetry and band extrema location. Herring and Vogt~\cite{herring56} formulated a more general acoustic deformation potential approach for cubic semiconductors with band extrema along several high-symmetry lines, which uses two or three independent deformation potential parameters. In spite of the successful applications of the Herring and Vogt model to cubic semiconductors~\cite{PhysRevB.78.035202, Cao2018, DSouza2020,Fischetti1996,Cao2020,DSouza2022},  the acoustic deformation potential model has not been laid out in sufficient detail for arbitrary crystal symmetries and band extrema locations.

Machine-learning techniques could provide another route to faster simulations of el-ph coupling and electronic transport in bulk and 2D semiconductors. There have been several reports of using machine learning (ML) to predict physical properties determined by el-ph coupling, such as zero-point renormalization~\cite{Haldar2024},  thermoelectric figure of merit~\cite{Bassman2018,Wang2020,Wang2023,Qu2024,Vaitesswar2024,Barua2024} and superconducting transition temperature~\cite{Stanev2018,Xie2022,Cerqueira2024, daSilva2025},
but without predicting el-ph matrix elements. \textcolor{black}{One of the approaches to obtain el-ph matrix elements using machine learning uses deep neural networks to learn the key quantity of DFPT i.e. the change of the Kohn-Sham
potential due to atomic and other perturbations~\cite{PhysRevLett.132.096401}. Another approach uses deep neural networks to construct atomic orbital-based Hamiltonian matrices and gradients due to atomic displacements trained on first-principles calculations, from which el-ph matrix elements are obtained~\cite{Zhong2024}. A recent work proposed a method to learn el-ph matrix elements from finite difference first-principles calculations using neural networks~\cite{arXiv:2602.23084}.} 
Another relevant work used the Gaussian process regression to average over different momentum directions and obtain average el-ph matrix elements to speed up the calculations of thermoelectric transport properties~\cite{Wee2019}. 
However, other ML approaches to calculate el-ph matrix elements may be possible. 
For example, relatively small sets of calculated el-ph matrix elements for each phonon mode could be used as the training dataset to predict the el-ph matrix elements on very dense grids in relevant parts of the Brillouin zone, but this has not been attempted in the literature so far.

In this paper, we present \textcolor{black}{computationally efficient} first-principles based model and machine-learning approaches to electron-phonon scattering  in semiconductors, \textcolor{black}{which rely on explicit calculations of much smaller sets of electron-phonon matrix elements compared to the DFPT+Wannier approach}. We use these methods to compute electronic thermoelectric transport properties of single layer MoS$_2$, which exhibits large thermoelectric power factors near 300 K.
Within the model approach, we generalize the Herring and Vogt acoustic deformation potential approach~\cite{herring56} for any crystal symmetry and band extrema location. We derive explicit formulas for hexagonal symmetry with conduction band valleys located along the $\Gamma$-K line in the Brilloin zone. We combine this model with the model of the Fr\"ohlich interaction for polar optical phonons in 2D semiconductors~\cite{sohier16,sio22} and the deformation potential models for other relevant phonons~\cite{ridley99} to obtain el-ph scattering rates and thermoelectric transport properties of 2D MoS$_2$.
\textcolor{black}{Within this framework, the electron–phonon interaction is described by a compact set of physically transparent parameters that can be reused in subsequent studies without repeating the full first-principles workflow e.g.~for device modeling.}
We also develop a machine-learning approach that uses regression methods to predict el-ph matrix elements on very dense grids in regions of the Brillouin zone that contribute to electronic transport using a small training set of el-ph matrix elements calculated {\it ab-initio} \textcolor{black}{(of the order of 100 matrix elements per electronic band and phonon mode)}. 
The formulation of the machine-learning method is very general and can be straightforwardly applied to materials of any symmetry. Both approaches are in good agreement with the full first-principles (DFPT+Wannier)  approach and experimental data for 2D MoS$_2$.

\section{Methodology}

\subsection{Thermoelectric transport properties and electron-phonon coupling}

The thermoelectric transport coefficients are calculated using the Boltzmann transport theory within the momentum relaxation time approximation. The thermoelectric transport kernel function ($L_\beta$), with $\beta$ denoting the order of the transport moment, can be expressed as \cite{ashcroft76} 
\begin{eqnarray}
L_{\beta} = \frac{2e^{2-\beta}}{V}\sum_{n,{\bm k}} \left(-\frac{\partial f_{n{\bm k}}^0}{\partial E_{n{\bm k}}}\right)\tau_{n\bm k} \overline{v}_{n\bm k}^2 (E_{n\bm k}-E_F)^{\beta}, \label{eq:trans_kernel}
\end{eqnarray}
where  $e$ is the electronic charge, $V$ is the crystal volume, $E_{n\bm k}$ is the energy of an electronic state with band index $n$ and crystal momentum ${\bm k}$, $\overline{v}_{n\bm k}$ is its average group velocity, $\tau_{n\bm k}$ is the momentum relaxation time, and $E_F$ is the Fermi energy. The electrical conductivity $\sigma$, the Seebeck coefficient $S$ and the power factor PF can then be written as
\begin{eqnarray}\label{eq: TE_para}
\sigma &=& L_0, \\
 S &=& \frac{L_1}{T L_0}, \\
 PF &=& S^2\sigma.
\end{eqnarray}
\textcolor{black}{The electron drift mobility $\mu$ (also referred to as the electron mobility from here on) is obtained from $\mu=\sigma/(en_c)$, where $n_c$ is the carrier concentration.}

The momentum relaxation times caused by electron-phonon interaction are calculated using Fermi's golden rule~\cite{Sohier2014,Gunst2016}
\begin{eqnarray}\label{eq:el-ph}
\tau_{n\bm k}^{-1} &=& \frac{2\pi}{\hbar} \textcolor{black}{\sum_{\pm} \sum_{\bm{k'},n'}} \sum_{\lambda,\bm{q}} |g_{n{\bm k},\lambda{\bm q}}^{n'{\bm k'}}|^2 \big(1-\cos\theta_{\bm{kk'}}\big) \\ \nonumber
&\times& \bigg[n(\omega_{\lambda{\bm q}})+\frac{1}{2} \mp \frac{1}{2} \bigg]
\frac{1-f^0(E_{n'\bm{k'}})}{1-f^0(E_{n\bm{k}})}\nonumber \\ \nonumber
&\times& \delta_{\bm{k'},\bm{k \pm q}}\delta(E_{n'\bm k'}-E_{n\bm k}\mp \hbar \omega_{\lambda{\bm q}}),
\end{eqnarray}
where $\hbar$ is the Planck constant and $g_{n{\bm k}\lambda{\bm q}}^{n'{\bm k'}}$ is the el-ph matrix element between electronic states $n{\bm k}$ and $n'{\bm k'}$ for a phonon mode $\lambda$ with the wave vector $\bm q$ and the phonon frequency $\omega_{\lambda{\bm q}}$. The angle between the wave vectors $\bm{k}$ and $\bm{k'}$ is denoted by $\theta_{\bm {kk'}}$, $f^0(E_{n\bm{k}})$ is the Fermi-Dirac distribution, and $n(\omega_{\lambda{\bm q}})$ is the Bose-Einstein distribution. The Kronecker symbol $\delta_{\bm{k'},\bm{k \pm q}}$ represents the crystal momentum conservation. 
The absorption (emission) of a phonon is represented by the upper (lower) sign.
The scattering rate for the electronic state $n\bm k$ can be obtained by excluding the terms $1-\cos\theta_{\bm{kk'}}$ and $[1-f^0(E_{n'\bm{k'}})]/[1-f^0(E_{n\bm{k}})]$ in Eq.~\eqref{eq:el-ph}. Details of the computational implementation of this approach can be found in Ref.~\cite{DSouza2020}.

\subsection{Model approach}

To develop an accurate model of electronic and thermoelectric transport in a material of interest, we need to analyze its  electronic and phonon band structure. Monolayer MoS$_2$ has a hexagonal symmetry and is a semiconductor with a direct band gap located at the K and K$'$ points of the Brillouin zone (see Fig.~\ref{fig: uc}). 
\textcolor{black}{The electronic band structure of monolayer MoS$_2$ is shown in Supplementary Material~\cite{supp} \nocite{Ando1982}.}
The higher conduction band minima (CBM) are at least 300 meV above the lowest CBM at K and K$'$, so the transport properties of $n$-type MoS$_2$ are governed only by the lowest conduction band states near K and K$'$. The energy dispersion near the CBM at K and K$'$ is nearly parabolic (Fig.~\ref{fig: uc}) and can be described using the Kane model~\cite{kane57}.  Since the lowest CBM are located at K and K$'$, we consider intravalley scattering mechanisms (within the K and K$'$ valleys) originating from the phonon modes near $\Gamma$ and intervalley scattering mechanisms (between the K and K$'$ valleys) originating from the phonon modes near K. 

\begin{figure}[h!]
\begin{centering}
\includegraphics[keepaspectratio, width=0.45\textwidth]{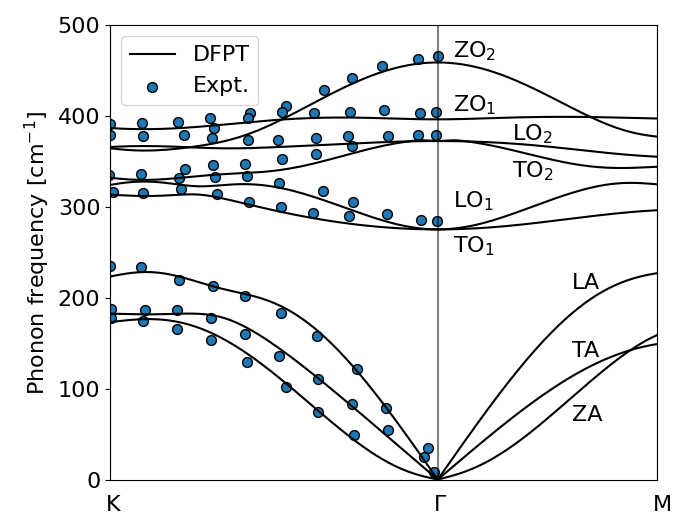}
\caption{\label{fig:ph} 
The phonon dispersion of single layer MoS$_2$ calculated using density functional perturbation theory, represented with solid lines. The experimental data taken from Ref.~\cite{Wakabayashi75} are shown with circles~\cite{mos2phonon}. The acoustic phonon branches are in-plane longitudinal (LA), in-plane transverse (TA) and out-of-plane (ZA). The optical modes are two in-plane longitudinal (LO) modes, two in-plane transverse (TO) modes, and two out-of-plane (ZO) modes. 
}
\end{centering}
\end{figure}

The phonon band structure of 2D MoS$_2$ is shown in Fig.~\ref{fig:ph}. The acoustic phonon branches are in-plane longitudinal (LA), in-plane transverse (TA) and out-of-plane (ZA). The optical modes are two in-plane longitudinal (LO) modes, two in-plane transverse (TO) modes, and two out-of-plane (ZO) modes. All the modes with the wave vectors corresponding to the $\Gamma$ 
point are either odd or even with respect to the in-plane mirror plane containing Mo atoms. The el-ph matrix elements corresponding to the intravalley 
scattering of the electronic states at K and K$'$ due to mirror-odd ZA, LO$_1$, TO$_1$, and ZO$_2$ modes at $\Gamma$ are zero by symmetry. Odd parity phonons can couple only electronic states of opposite parity since el-ph matrix elements are invariant under symmetry operations~\cite{dresselhaus08}. The el-ph matrix elements due to the mirror-odd phonons at $\Gamma$ vanish exactly at the K and K$^{'}$ points since the initial and final electronic states are of the same parity. We therefore account for only mirror-even LA, TA, LO$_2$, TO$_2$, and ZO$_1$ modes in our treatment of intravalley scattering. 
 
\begin{figure}[h!]
\begin{centering}
\includegraphics[keepaspectratio, width=0.45\textwidth]{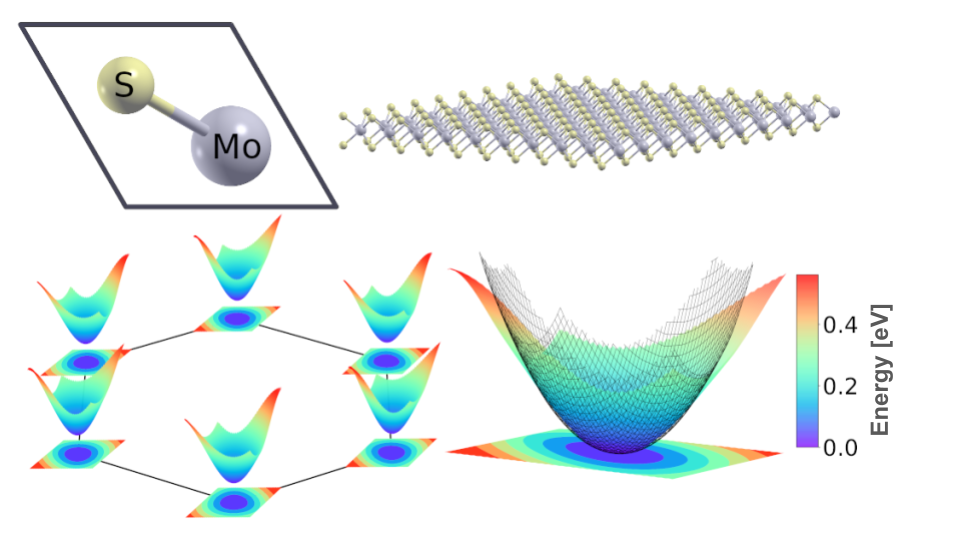}
\caption{\label{fig: uc}
(Top left) Top view of the hexagonal unit cell of monolayer MoS$_2$. (Top right) Atomic structure of a MoS$_2$ sheet. (Lower left) The energy dispersion of the lowest conduction band of MoS$_2$, calculated using density functional theory (DFT) in the vicinity of the K and K$'$ points in the hexagonal Brillouin zone. (Lower right) The black curve is the dispersion of the conduction band obtained using the Kane model fitted to DFT results, shown in color. }
\end{centering}
\end{figure}

We describe the intravalley interaction of the electronic states near K and K$'$ with LA and TA modes using the acoustic deformation potential approach~\cite{herring56}, which we generalize to arbitrary crystal symmetries and band extrema locations. The intravalley scattering with the non-polar 
optical mode (TO$_2$ 
near $\Gamma$)
and the intervalley scattering with zone-edge phonons near K
are described using the 
non-polar optical/intervalley 
deformation potential approach~\cite{ridley99}. 
The non-polar optical deformation potential approach was modified to empirically include the effect of long-range interactions on the intravalley scattering with homopolar optical modes (ZO$_1$ near $\Gamma$). 
 We use the Fr\"ohlich-like model of the long-range dipole-dipole interaction to describe scattering due to  the polar optical mode (LO$_2$ near $\Gamma)$~\cite{sohier16,sio22}.  We do not account for the   the long-range piezoelectric interaction associated with acoustic phonons which is limited to low temperatures~\cite{Kaasbjerg2013}. 
Charge screening effects on electron-phonon interaction
 are also not included in our analysis.

\subsubsection{Intravalley acoustic scattering: Generalized deformation potential model}

The interaction between electrons and acoustic phonons in the long-wavelength limit is characterized by a slowly varying potential, which can be expressed as \cite{herring56,ridley99,yu05}
\begin{eqnarray}\label{eq:Hep}
H_{ep} = \sum_{\alpha\beta}\Xi_{\alpha\beta}\epsilon_{\alpha\beta},
\end{eqnarray}
where $\Xi$ is the deformation potential (DP) tensor, $\epsilon$ is the strain tensor, and $\alpha$ and $\beta$ are the Cartesian coordinates. A second-rank symmetric tensor has a maximum of six independent components. However, depending on the material's symmetry and the position of the electronic band extrema in the Brillouin zone, symmetry constraints reduce the number of independent components in the DP tensor. In Supplementary Material~\cite{supp}, we outline a procedure to obtain the number of independent DP components for a crystal of any symmetry and any position of the electronic band extrema. Our method represents a generalization of the Herring and Vogt DP approach for cubic crystals whose band extrema lie along the $\Gamma$-L, $\Gamma$-X and $\Gamma$-K directions~\cite{herring56}. 

\begin{center}
\begin{table}[h]
\caption{Expressions for the six components of the deformation potential tensor for a bulk crystal with hexagonal symmetry and band extrema along the $\Gamma$-K direction.}\label{tab:sym_res}
\renewcommand{\arraystretch}{1.5}
\begin{tabular}{ c | c }
\hline
Valley direction & $\Gamma$-K \\
\hline
\hline
$\Xi_{1}$ = $\Xi_{xx}$ & $\Xi_d$ + $\Xi_u$ \\ 
$\Xi_{2}$ = $\Xi_{yy}$ & $\Xi_d$ + $\Xi_u$ - $\Xi_p$  \\ 
$\Xi_{3}$ = $\Xi_{zz}$ & $\Xi_d$ - $\Xi_u$ + $\Xi_p$  \\ 
$\Xi_{4}$ = $\Xi_{yz}$ = $\Xi_{zy}$ & 0  \\ 
$\Xi_{5}$ = $\Xi_{xz}$ = $\Xi_{zx}$ & 0  \\ 
$\Xi_{6}$ = $\Xi_{xy}$ = $\Xi_{yx}$ & 0 \\ 
\end{tabular}
\end{table}
\end{center}

For a three-dimensional (3D) unit cell with hexagonal symmetry, symmetry constrains dictate that the DP tensor elements can be characterized in terms of three independent components when the band extrema are along the $\Gamma$-K direction (see Supplementary Material~\cite{supp}). These components have direct physical interpretation. Using the Herring and Vogt definitions of volume dilation ($\Xi_d$), uniaxial stress ($\Xi_u$) and shear ($\Xi_p$) \cite{supp}, we can express the DP tensor for a 3D material with hexagonal symmetry and band extrema along the $\Gamma$-K direction as shown in Table~\ref{tab:sym_res}. In the case of a 2D hexagonal crystal, the $\Xi_3$ component of the DP tensor is zero, and the number of independent components reduces to two ($\Xi_d$ + $\Xi_u$ and $\Xi_p$). \textcolor{black}
{The procedure to calculate these deformation potentials is outlined in Sec.~\ref{sec:parameters} and Appendix~\ref{ap: in_ac_dp}.}

We obtain electron-phonon matrix elements due to scattering with long-wavelength in-plane acoustic modes using the DP tensor described above. Since the acoustic phonon frequencies near the zone center are small, we use $n(\omega_{\lambda{\bm q}}) \approx \frac{k_BT}{\hbar \omega_{\lambda{\bm q}}}$ for both absorption and emission processes, where $k_B$ is the Boltzmann constant, $T$ is the temperature, and $|{\bm q}|\rightarrow 0$. At room temperature, the equipartition between absorption and emission processes holds, {\it i.e.} $n(\omega_{\lambda,{\bm q}}) \gg 1/2$. Therefore, if we neglect acoustic phonon energies when calculating the scattering rate given by Eq.~\eqref{eq:el-ph}, the scattering rates due to emission and absorption are approximately the same. 
If we assume that phonon frequencies change linearly with ${\bm q}$ in the long-wavelength limit, 
we obtain \cite{herring56,DSouza2020}
\begin{eqnarray}\label{eq:ac} 
|M_{\lambda{\bm q}}|^2 \textcolor{black}{\equiv} |g_{\lambda{\bm q}}|^2 \times 2 n(\omega_{\lambda{\bm q}})\nonumber\\=\frac{k_B T I^2({\bm{k,k'}})}{A}\frac{\sum_{ij}\Xi_i \Xi_j f_i f_j}{\sum_{ij} c_{ij} f_i f_j}.
\end{eqnarray}
Here $g_{\lambda{\bm q}}$ is the el-ph matrix element between the lowest conduction band states with wave vectors ${\bm k}$ and ${\bm k'}$ and $ I({\bm k},{\bm k'})$ is the overlap integral between the two Bloch states with the wave vectors ${\bm k}$ and ${\bm k'}$. $A$ is the unit cell area, $c_{ij}$ are the components of the 6 $\times$ 6 elastic constant tensor, and $f_i$ are the components of the $3 \times 3$ polarization tensor given as
\begin{eqnarray}\label{eq:pol}
\begin{gathered}
f_i = a_iq_i, \; i = 1,2,3,\\
f_4 = a_2q_3 + a_3q_2, \\
f_5 = a_1q_3 + a_3q_1, \\
f_6 = a_1q_2 + a_2q_1,\\
\end{gathered}
\end{eqnarray}
where ${\bm a}$ is the atomic displacement for an acoustic phonon with wave vector ${\bm q}$. We calculate the polarization vectors by solving the generalized equation of motion for the atomic displacement in a hexagonal crystal~\cite{ridley99}
\begin{widetext}
\begin{align}\label{mat:ang_dep}
\begin{pmatrix}
\vspace{0.1cm}
c_{11}\alpha^2 + \frac{c_{11}-c_{12}}{2}\beta^2 + c_{44}\gamma^2-\rho v_s^2 & \frac{c_{11}+c_{12}}{2}\alpha\beta & (c_{13}+c_{44})\alpha\gamma \\\vspace{0.1cm}
\frac{c_{11}+c_{12}}{2}\alpha\beta & c_{11}\beta^2 + \frac{c_{11}-c_{12}}{2}\alpha^2 + c_{44}\gamma^2 -\rho v_s^2 & (c_{13}+c_{44})\beta\gamma \\\vspace{0.1cm}
(c_{13}+c_{44})\alpha\gamma & (c_{13}+c_{44})\beta\gamma & c_{44}(\alpha^2 + \beta^2) + c_{33}\gamma^2 -\rho v_s^2 \\
\end{pmatrix}
\begin{pmatrix}
a_1 \\
a_2 \\
a_3
\end{pmatrix}
= 0,
\end{align}
\end{widetext}
where $\rho$ is the mass density, and $v_s$ is the speed of sound. The direction cosines of the wave vector $\bm q$ are denoted by $\alpha$, $\beta$ and $\gamma$. In the case of a 2D hexagonal crystal, only $c_{11}$ and $c_{12}$ are not zero. We can obtain the matrix elements for the in-plane longitudinal and transverse acoustic branches along any ${\bm q}$ direction  by substituting the polarization tensor components derived from Eq.~\eqref{mat:ang_dep} and the DP components from Table~\ref{tab:sym_res} in Eq.~\eqref{eq:ac}.
The expressions for $\sum_{ij}\Xi_i \Xi_j f_i f_j/\sum_{ij} c_{ij} f_i f_j$ [see Eq.~\eqref{eq:ac}] along the high symmetry directions are listed in Table~\ref{tab:LaTa-cont}.

\begin{table}[h!]
\caption{Longitudinal acoustic (LA) and transverse acoustic (TA) mode contributions to the square of the electron-phonon matrix elements for the K valley [i.e. $\sum_{ij}\Xi_i \Xi_j f_i f_j/\sum_{ij} c_{ij} f_i f_j$ in Eq.~\eqref{eq:ac}] along the high symmetry directions.  $\Xi$ and $c$ are the 6$\times$6 deformation potential and elastic constant tensors, respectively. ${\bm q}$ is the phonon wave vector.
} 
\label{tab:LaTa-cont} 
\renewcommand{\arraystretch}{2.5}
\begin{tabular}{c|c|c|c}
Valley & ${\bm q}$ direction  & LA & TA \\ \hline \hline
K & $\parallel$ to $\Gamma$-K & $\frac{(\Xi_d + \Xi_u)^2}{c_{11}}$ & 0 \\ \hline
\ &$\parallel$ to $\Gamma$-M & 
$\frac{(\Xi_d + \Xi_u - \frac{1}{4}\Xi_p)^2}{c_{11}}$
& $\frac{3\Xi_p^2}{8(c_{11} - c_{12})}$  \\ 
\end{tabular}
\end{table}

\subsubsection{Intravalley non-polar optical phonon and intervalley scattering}
\label{sec:non-polar}

Intervalley scattering between the K and K$'$ valleys occurs via
phonons near the K point. Scattering via phonons near the Brillouin zone edge typically has a weak wave vector dependence~\cite{ridley99}. The el-ph matrix element between the lowest conduction band states with wave vectors \textcolor{black}{${\bm k}\approx$ K and ${\bm k'}\approx$ K$'$ can therefore be approximated as}
\begin{equation}\label{eq:inter}
g_{\lambda{\bm q}}  \approx g_{\lambda{\bm Q}} =  \sqrt{\frac{\hbar}{2m \omega_{\lambda{\bm Q}}}}\Xi_{\lambda{\bm Q}},
\end{equation}
\textcolor{black}{where ${\bm Q}$ corresponds to the difference between the wave vectors of the K and K$'$ points, $m$ is the mass of the atoms in the unit cell and $\Xi_{\lambda{\bm Q}}$ is the deformation potential due to scattering between the valleys K and K$'$ via the phonon mode $\lambda$.} 
We also approximate the frequencies of zone edge phonons to be equal to $\omega_{\lambda{\bm Q}}$ when calculating the relaxation times using Eq.~(\ref{eq:el-ph}). These approximations are also valid for intravalley scattering due to short wave vector non-polar optical phonons~\cite{ridley99} \textcolor{black}{i.e. we  use Eq.~(\ref{eq:inter}) with ${\bm Q}={\bm 0}$ to approximate the el-ph matrix elements with ${\bm k}\approx $ K and  ${\bm k'}\approx $ K. The calculation of all these deformation potentials is given in Sec.~\ref{sec:parameters}.} 

\subsubsection{Intravalley homopolar optical phonon scattering}
It has been recently established that the el-ph matrix elements associated with the homopolar optical ZO$_1$ mode near $\Gamma$ have a small long-range contribution~\cite{Deng2021,Ponce2023_PRB,Ponce2023_PRL} 
Therefore, these el-ph matrix elements exhibit stronger wave vector dependence compared to the zone-center non-polar optical TO$_2$ mode.
We parametrize the scattering due to the homopolar optical mode near $\Gamma$ using the following form of the deformation potential

\begin{eqnarray}\label{eq:hp_q}
\Xi^{ZO_1} &=& \sum_{i =0}^4 \Xi^{ZO_1}_{i} q^i. 
\end{eqnarray}
The parameters $\Xi^{ZO_1}_i$ ($i = 0, 4 $) are the Taylor series coefficients of the homopolar optical deformation potential expanded in powers of the phonon wave vector magnitude $q$. \textcolor{black}{Sec.~\ref{sec:parameters} describes the procedure to calculate these coefficients.}

\subsubsection{Intravalley polar optical phonon scattering}

In addition to the deformation potential in the long-wavelength limit, the atomic displacements corresponding to the LO$_2$ mode of MoS$_2$ create long-range interactions between electrons and LO$_2$ modes. This long-range interaction can be described using the expression for the Fr\"ohlich interaction in 2D crystals~\cite{sohier16,sio22}
\begin{multline}\label{eq:fr}
g_{\lambda{\bm q}} = i \Big[\frac{\pi}{2}\frac{e^2}{4\pi\varepsilon_0}\frac{d}{A}\hbar \omega_{LO_2} (\epsilon_0 - \einf)\Big]^{\frac{1}{2}} \times \\
\frac{1}{\einf}\frac{2}{q d}\Big[1+(\einf q d)^{-1} \frac{\expqd-1}{1-(1+\einf^{-1})(1+\expqd)/2}\Big].
\end{multline}
where $\omega_{LO_2}$ is the frequency of the LO$_2$ phonon at $\Gamma$, $d$ is the effective dielectric thickness, $\epsilon_{\infty}$ is the high-frequency dielectric constant of the monolayer, and $\epsilon_0$ is the dielectric constant of vacuum. \textcolor{black}
{The procedure to calculate these parameters is outlined in Supplementary Material~\cite{supp}.} 

\subsection{Machine-learning approach}
\label{sec:ML}
Our machine-learning method to compute el-ph matrix elements relevant for thermoelectric transport in semiconductors is based on regression approaches, which are used to predict el-ph matrix elements near band extrema on a very dense grid using a relatively small training dataset calculated using the DFPT+Wannier approach. Our method is straightforward because it is possible to fairly accurately describe those el-ph matrix elements using the analytical formulas of the model approach.

The ML approach to calculate el-ph matrix elements near band extrema from a training dataset is carried out using the  `{\sc{Scikit learn}}' package \cite{scikit-learn}. The training datasets can be chosen in different ways. We tested the training dataset consisting of first-principles el-ph matrix elements \textcolor{black}{where the initial electronic state ${\bm k}$ is located at the K point and the phonon wave vectors ${\bm q}$} are uniformly distributed along selected directions in the first Brillouin zone, which include high-symmetry directions [as sketched in the inset of  Fig. \ref{fig:train_ML} (a)]. 
We will refer to this type of training datasets as "uniform".
We also used the training datasets consisting of \textcolor{black}{${\bm k}=$ K} and randomly chosen 
\textcolor{black}{${\bm q}$-points} in all directions of the first Brillouin zone. In both datasets, the $\Gamma$ point is included \textcolor{black}{in the ${\bm q}$-point set} and the magnitude of the phonon wave vectors 
varies between 0 and 35\% of the distance between the $\Gamma$ and K points. We sample only this small part of the Brillouin zone as only the phonon wave vectors near $\Gamma$ contribute significantly to intravalley scattering.
We do not use machine learning for intravalley non-polar and intervalley scattering since they are well described by only one el-ph matrix element (as described in Sec.~\ref{sec:non-polar}). In this case, we explicitly calculate that matrix element using DFPT and then use analytical formulas of Sec.~\ref{sec:non-polar} to compute the scattering.

For intravalley scattering due to acoustic modes (LA and TA), we use the ``{\it PolynomialFeatures}" (PFR) algorithm to obtain el-ph matrix elements near band extrema since the acoustic deformation potential model give the polynomial (linear) dependence on the phonon wave vector. The PFR approach generates a matrix consisting of all possible combinations of polynomials with degrees less than or equal to the user stated input value. The specified degree for all acoustic modes were set at 5.

For intravalley scattering due to the polar and homopolar optical modes (LO$_2$ and ZO$_1$), the ``{\it GaussianProcessRegressor}" (GPR) algorithm is used to compute el-ph matrix elements near band extrema. This is done because the dependence of these matrix elements on the phonon wave vector is more complicated than that for intravalley acoustic scattering, as can be seen from our model approach. GPR uses the Gaussian process for regression and hence a prior mean and covariance must be defined.
We use the mean of the training set data as the prior mean.
The prior's covariance is obtained by implementing a kernel object.
We implement the radial-basis function (RBF) kernel (or ``squared-exponential kernel"). 
The RBF kernel defined on two feature vectors, labeled as $x$ and $x'$, is expressed as 
\begin{eqnarray}
K(x,x')= \exp(\frac{-|d(x,x')|^2}{2l^2}),
\end{eqnarray}
where the hyperparameters $d(x,x')$ and $l$ are the Euclidean distance and the length scale of the kernel, respectively.  These hyperparameters are then obtained by maximizing the log-marginal likelihood.

The el-ph matrix elements obtained as described above are then used in Eq.~\eqref{eq:el-ph} to compute el-ph relaxation times and scattering rates. At present, we approximate the electronic band structure with the nearly parabolic (Kane) model~\cite{kane57} and treat the frequencies of the phonon modes \textcolor{black}{as in our model approach (acoustic phonon frequencies depend linearly on $q$, while all other phonon frequencies are independent on the phonon wave vector and their values are taken at $\Gamma$ or K).} The electronic and phonon band structures could also be interpolated using a similar ML approach as described here, which may be addressed in future work. 

\subsection{First principles calculations}

The DFT calculations of the electronic band structure of single layer MoS$_2$ are carried out using the {\sc{quantum espresso}} code \cite{qe09,qe17}. We used a 24 $\times$ 24 $\times$ 1 reciprocal space $\bm{k}$-point grid for plane waves and the Perdew-Burke-Ernzerhof (PBE) exchange-correlation functional with spin-orbit coupling (SOC) \cite{pbe96}.
The kinetic and charge density energy cut-off used in our DFT calculations are 70 Ry and 280 Ry, respectively.
We utilize the 2D Coulomb cutoff approach to avoid interaction between MoS$_2$ layers \cite{sohier17}.
The vacuum spacing is 22~\AA\ along the direction perpendicular to the plane of the MoS$_2$ layer. 

We compute the phonon dispersion of single layer MoS$_2$  using  density functional perturbation theory (DFPT) and the {\sc{quantum espresso}} code. An 12 $\times$ 12 $\times$ 1 phonon wave vector $\bm{q}$-point grid is used in this calculation. The el-ph matrix elements are calculated on the coarse 24 $\times$ 24 $\times$ 1 electron wave vector $\bm{k}$- and 12 $\times$ 12 $\times$ 1 phonon wave vector $\bm{q}$-point grids. 
We then use the Wannier interpolation approach, as implemented in the {\sc EPW} code~\cite{Lee2023,Ponce2016}, to
interpolate these matrix elements onto dense 204 $\times$ 204 $\times$ 1 $\bm{k}$- and $\bm{q}$-point grids by contriving 22 Wannier orbitals.
The initial guess are the $d$ orbitals for the Mo atoms and the $p$ orbitals for the S atoms which are inclusive of both spin-up and spin-down orbitals. The values of quadripoles for the accurate treatment of long-range interactions are taken from Ref.~\cite{Ponce2023_PRL}.

\section{Results and discussion}

\subsection{First-principles parameters for the model approach}
\label{sec:parameters}

The energy dispersion near the CBM at K and K$'$ can be expressed with the Kane Hamiltonian \cite{kane57}
\begin{eqnarray}\label{eq:kane_L}
\frac{\hbar^2k^2}{2m^*} = E(1+\alpha_p E), 
\end{eqnarray}
where $\hbar$ is the Planck constant, $m^*$ is the effective mass, $k$ is the magnitude of the wave vector with respect to the CBM, $E$ is the electronic energy with respect to the CBM and $\alpha_p$ is the measure of the deviation from the parabolic dispersion relation. The effective mass and the non-parabolic factor are calculated by fitting Eq.~\eqref{eq:kane_L} to the conduction band energy dispersion obtained from DFT, as shown in Fig.~\ref{fig: uc}.  Our calculated effective masses and the non-parabolic factor are shown in Table~\ref{tab:lat_para}, along with structural parameters.
\textcolor{black}{
The calculated lattice constant, the monolayer thickness and the band gap
 are nearly identical to those obtained in a recent study at the PBE level of theory (less than 0.1\% difference)~\cite{ Ponce2023_PRB}, while there is a 3.7\% difference for the effective mass. To the best of our knowledge, no literature values are available for the non-parabolicity parameter $\alpha_p$.}

\begin{table}[h!]
\caption{\label{tab:lat_para} Calculated lattice constant $a_0$, the monolayer thickness $h_z$, the band gap $E_g$,  the effective mass $m^*$ and the non-parabolic factor $\alpha_p$ of monolayer MoS$_2$.}
\renewcommand{\arraystretch}{1.5}
\begin{tabular}{c|c|c}
Parameters & Calculation & Experiment \\
\hline
\hline
$a_0$ [\AA] & 3.188 & 3.16  \hyperlink{mylink1}{$^{a}$} \\
$h_z$ [\AA] & 3.133 & \  \\
$E_g$ [eV] & 1.6 & 1.88 \hyperlink{mylink2}{$^{b}$} \\
$m^*$ [$m_e$] & \textcolor{black}{0.405} & \\ 
$\alpha_p$ & 0.7 & \  \\
\hline
\end{tabular}
\begin{tablenotes}
\item \hypertarget{mylink1}{ $^a$ Ref. \cite{Blue2020}}
\item \hypertarget{mylink2}{ $^b$ Ref. \cite{mak10}}
\end{tablenotes}
\end{table}

The parameters describing el-ph coupling were obtained as follows. The independent components of the DP tensor for LA and TA modes were computed from the calculations of \textcolor{black}{$\approx 10$} el-ph matrix elements in the limit of small phonon wave vectors along selected high-symmetry lines using DFPT calculations and the Wannier interpolation, as described in Appendix \ref{ap: in_ac_dp}. Our computed values of $\Xi_d$ + $\Xi_u$ and $\Xi_p$ are given in Table \ref{tab:dp}, together with the values of elastic constants \textcolor{black}{computed using DFPT and a finite-difference method.}
\textcolor{black}{Our elastic constant values are in excellent agreement with previously reported PBE values \cite{akr2014}.}
The deformation potentials for the homopolar optical mode were obtained in the similar manner as acoustic deformation potentials. They were fitted along the $\Gamma$-K direction, and it was checked that the same fits were obtained along the $\Gamma$-M direction.  
The deformation potentials for intervalley and intravalley non-polar optical phonon scattering were calculated using Eq.~\eqref{eq:inter}, where the el-ph matrix elements \textcolor{black}{$g_{\lambda{\bf Q}}$} were obtained using DFPT for all relevant phonons with the wave vectors corresponding to the $\Gamma$ and K points. Their values and the corresponding phonon frequencies are listed in Supplementary Material \cite{supp}. \textcolor{black}{Our deformation potential values are of the same order of magnitude as reported in Refs.~\cite{Kaasbjerg2012,Jin2014}, which used the local density approximation and supercell/DFPT methods to calculate el-ph matrix elements without an accurate treatment of long-range interactions. Our el-ph matrix elements agree well with the state-of-the-art DFPT+Wannier results obtained using PBE and accurately treating long-range interactions \cite{Ponce2023_PRB, Ponce2023_PRL}.} \textcolor{black}{The computational cost of extracting deformation potential parameters is discussed in Supplementary Material~\cite{supp}.}
Finally, we computed the values of the effective dielectric thickness $d$ and the high-frequency dielectric constant $\epsilon_{\infty}$ that quantify the 2D Fr\"ohlich interaction using the capacitor stack model (see Supplementary Material \cite{supp}). 
These values are given in Table \ref{tab:dp} \textcolor{black}{ and are in excellent agreement with those reported in Ref.~\cite{sio22}, with deviations of approximately 0.2\% for $d$ and 2\% for $\epsilon_{\infty}$.}

\begin{center}
\begin{table}[h]
\caption{Parameters related to electron-phonon interaction used in our model approach (see the text for definitions), calculated from first principles.}
\renewcommand{\arraystretch}{1.5}\label{tab:dp}
\begin{tabular}{ c  c }
\hline
Intravalley & Parameters \\
\hline
In-plane acoustic modes & \\
\hline
$\Xi_{d} + \Xi_{u}$ [eV] & 7.20 \\ 
$\Xi_{p}$ [eV] & 2.63  \\
$c_{11}$[N/m] & \textcolor{black}{132.7}  \\
$c_{12}$[N/m] & \textcolor{black}{33} \\
\hline
Homopolar mode & \\
\hline
$\omega_{ZO_1}$ [meV] & 49.17 \\
$\Xi^{ZO_1}_{0}$ [$\frac{eV}{\AA}$] & 2.89 \\
$\Xi^{ZO_1}_{1}$ [eV] & -12.58 \\
$\Xi^{ZO_1}_{2}$ [eV $\cdot \AA$] & 28.65 \\
$\Xi^{ZO_1}_{3}$ [eV $\cdot \AA^2$] & -107.61 \\
$\Xi^{ZO_1}_{4}$ [eV $\cdot \AA^3$] & 184.96 \\
\hline
\hline
Polar mode & \\
\hline
 $\omega_{LO_2}$ [meV] & 46.2 \\
$d$ [\AA] & 5.48\\
$\einf$ & 16.76\\
\hline
\end{tabular}
\end{table}
\end{center}

\begin{figure*}[!]
\begin{centering}
\includegraphics[keepaspectratio, width=0.95\textwidth]{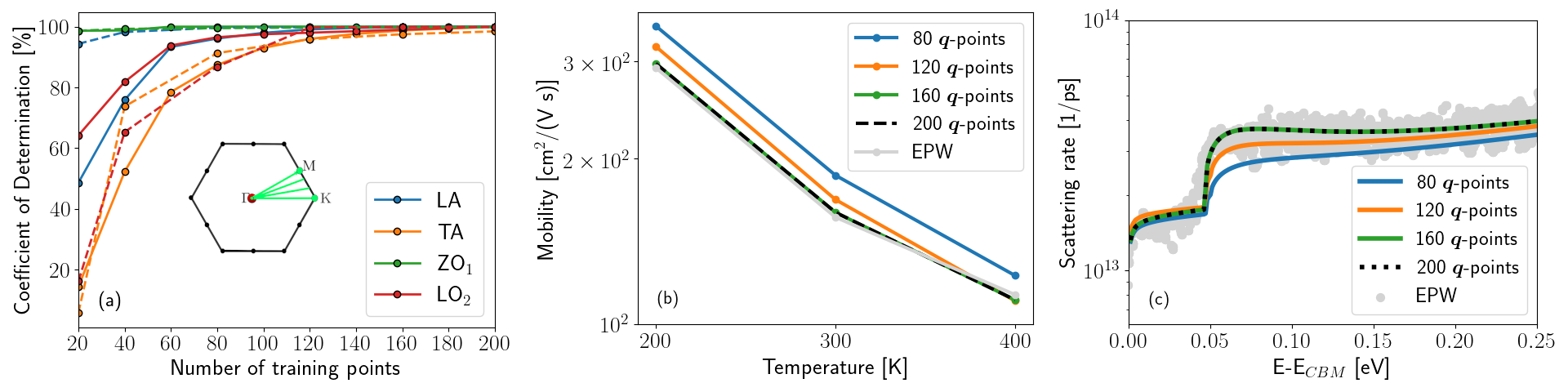}
\caption{\label{fig:train_ML} (a) Coefficient of determination for phonon-mode resolved electron-phonon matrix elements as a function of the size of the training dataset.  The results corresponding to the datasets consisting of \textcolor{black}{1 ${\bm k}$-point located at K and uniformly distributed ${\bm q}$-points} along selected lines in the Brillouin zone (the "uniform" datasets) are shown by solid lines. The results obtained from the datasets with \textcolor{black}{1 ${\bm k}$-point located at K and randomly chosen ${\bm q}$-points} are shown by dashed lines. LA and TA correspond to longitudinal and transverse acoustic phonons, respectively. ZO$_1$ and LO$_2$ stand for the homopolar and polar optical phonons, respectively. Inset: Selected directions in the first Brillouin zone along which the training dataset of first-principles electron–phonon matrix elements was uniformly distributed. (b) Electron \textcolor{black}{drift} mobility versus temperature and (c) scattering rate versus energy with respect to the conduction band minima at K and K$'$, obtained using the "uniform" training dataset for different dataset sizes, as well as the EPW code.} 
\end{centering}
\end{figure*}

\subsection{Training datasets for the machine-learning approach}

We verify the accuracy of the ML approach using the "uniform" and random training datasets explained in Sec.~\ref{sec:ML}. 
For this, we compute the coefficient of determination for the el-ph matrix elements 
obtained from regression compared to the those obtained using the DFPT+Wannier approach and the EPW code. These matrix elements were calculated for \textcolor{black}{the initial electronic state ${\bm k}$ located at K and the phonon wave vectors ${\bm q}$ located in} 
the circle centered at the $\Gamma$ point with the radius of 35\% of the distance between the $\Gamma$ and K points. Fig.~\ref{fig:train_ML}~(a) shows those coefficients of determination for the "uniform" and random datasets as a function of the size of the training set. We can see that the results obtained using the random datasets have a higher accuracy for somewhat smaller datasets compared to those obtained from the "uniform" datasets, except in the case of the polar optical phonon.
Both types of training datasets yield all the coefficients of determination above 90\% for the dataset sizes of $\sim$100. 

Fig.~\ref{fig:train_ML} (b) shows the electron \textcolor{black}{drift} mobility as a function of temperature for the "uniform" datasets with different sizes, as well as that calculated using the EPW code, \textcolor{black}{which computes the mobility using a linearized Boltzmann transport equation.} This figure  shows that the very accurate results for the mobility, compared to the state-of-the-art DFPT+Wannier approach, can be obtained with the "uniform" training datasets containing more than 100 ${\bm q}$-points \textcolor{black}{and converging for the datasets containing 160 ${\bm q}$-points.} Fig.~\ref{fig:train_ML} (c) illustrates the same point by showing the total scattering rate as a function of the energy with respect to the CBM, obtained using our ML approach and the DFPT+Wannier approach. 
For the results obtained using the ML approach from now on, we use the "uniform" training dataset with 160 ${\bm q}$-points.
\textcolor{black}{The computational cost of the ML approach is discussed in Supplementary Material~\cite{supp}.}

\begin{figure*}[!]
\begin{centering}
\includegraphics[keepaspectratio, width=\textwidth]{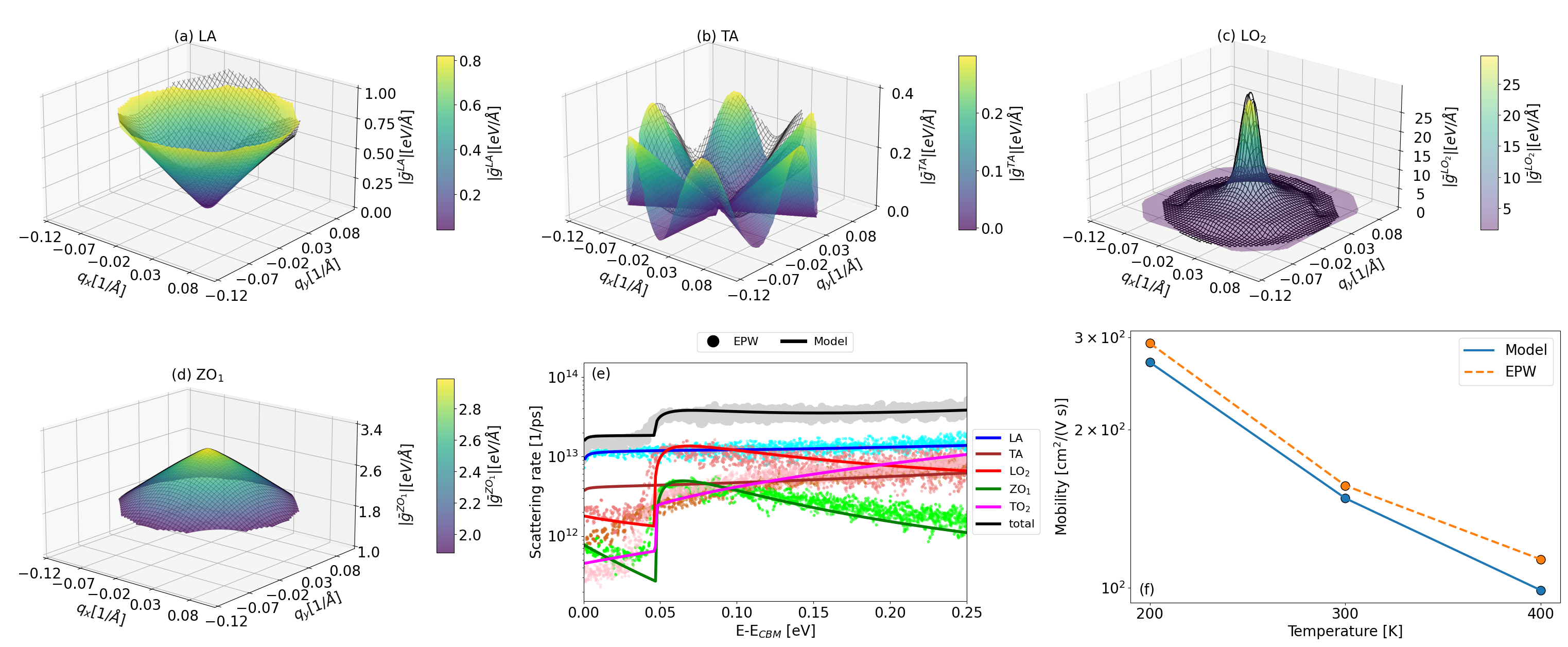}
\caption{\label{fig:dp} 
Phonon-mode resolved scaled electron-phonon matrix elements with the initial electronic state at the K point 
as a function of the phonon wave vector. The results shown in color are obtained using the model approach. The black curves are the results obtained using the EPW code. Panels (a), (b), (c) and (d) correspond to longitudinal acoustic (LA), transverse acoustic (TA), polar optical (LO$_2$) and homopolar optical (ZO$_1$) phonons, respectively. (e) Scattering rates  versus energy with respect to the conduction band minima at K and K$'$. Solid lines represent the model results, while the dots correspond to the EPW results. (f) Electron mobility versus temperature. Solid lines represent the model results, while dashed lines correspond to the EPW results.
}
\end{centering}
\end{figure*}

\subsection{Accuracy of the model and machine-learning approaches}

\begin{figure*}[t!]
\begin{centering}
\includegraphics[keepaspectratio, width=\textwidth]{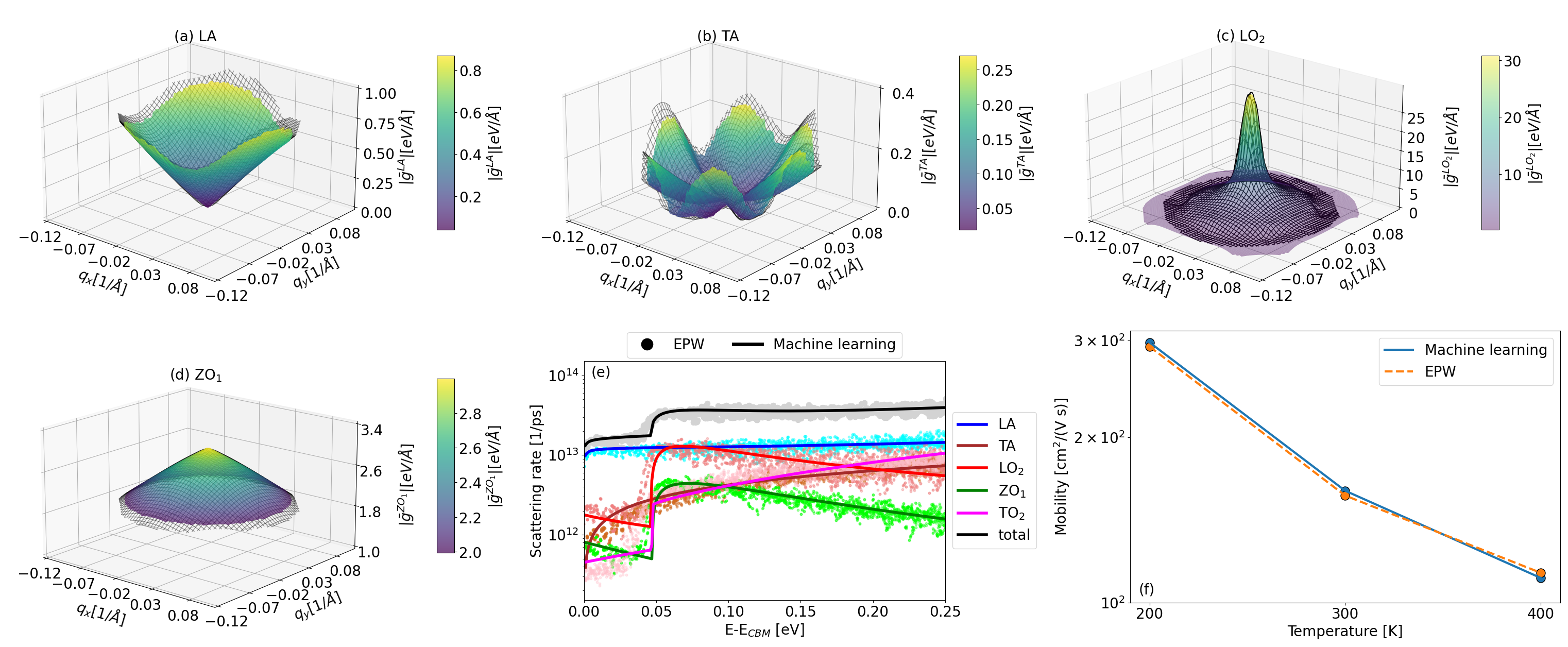}
\caption{\label{fig:ML} 
Phonon-mode resolved scaled electron-phonon matrix elements with the initial electronic state at the K point as a function of the phonon wave vector. The results shown in color are obtained using the machine-learning approach. The black curves are the results obtained using the EPW code. Panels (a), (b), (c) and (d) correspond to longitudinal acoustic (LA), transverse acoustic (TA), polar optical (LO$_2$) and homopolar optical (ZO$_1$) phonons, respectively.
 (e) Scattering rates  versus energy with respect to the conduction band minima at K and K$'$. Solid lines represent the machine-learning results, while the dots correspond to the EPW results. (f) Electron mobility versus temperature. Solid lines represent the machine-learning results, while dashed lines correspond to the EPW results.}
\end{centering}
\end{figure*}

We first compare the el-ph matrix elements obtained using our model and ML approaches with those obtained from DFPT and the Wannier interpolation computed with the EPW code. 
In Figs.~\ref{fig:dp} and \ref{fig:ML} (a)-(d), we plot the scaled el-ph matrix elements $
\bar{g}_{\lambda{\bm q}}  = g_{\lambda{\bm q}}/\sqrt{\hbar/2m \omega_{\lambda{\bm q}}}$ (for an easy extraction of deformation potentials, see Eq.~\eqref{eq:inter} and Ref. \cite{Murphy2018}) arising from the interaction with LA, TA, polar optical and homopolar optical phonons, respectively, where the K point is the initial electronic state. %
The EPW results are shown in \textcolor{black}{black} in both figures, while the results from the model and ML approach are shown in \textcolor{black}{color}.
It is evident that both the model and ML el-ph matrix elements match those of the DFPT+Wannier approach well for all phonon modes. 

The model and ML el-ph matrix elements also accurately capture the phonon wave vector dependence of the el-ph matrix elements obtained using the EPW code, see Figs. \ref{fig:dp} and \ref{fig:ML}. In all these approaches, the el-ph matrix elements corresponding to LA, LO$_2$ and ZO$_1$ phonons are isotropic, while those related to TA phonons are highly anisotropic~\cite{Kaasbjerg12}. The el-ph matrix elements due to the polar optical (LO$_2$) phonon mode steeply increase towards the $\Gamma$ point, but do not diverge~\cite{sohier16,sio22}. The el-ph matrix elements due to the homopolar (ZO$_1$) mode exhibit a cusp near the $\Gamma$ point due to the long-range interaction~\cite{Deng2021,Ponce2023_PRL}.
Both the model and ML approaches correctly describe the different behavior of el-ph matrix elements with respect to the phonon wave vector for all phonon modes.

Next we analyze the phonon-resolved scattering rates of single layer MoS$_2$ obtained from our approaches and the EPW code. 
In Figs.~\ref{fig:dp} and \ref{fig:ML} (e), we plot the scattering rates as a function of the electronic energy with respect to the CBM for the phonon modes that are mirror-even at $\Gamma$ (the scattering due to the phonon modes that are mirror-odd at $\Gamma$ is much weaker). The circular points correspond to the EPW calculations while the solid lines correspond to our model approach [Fig.~\ref{fig:dp} (e)] and our ML approach [Fig.~\ref{fig:ML} (e)].
We can see that the ML approach is almost as accurate as the DFPT+Wannier approach, while the model becomes somewhat less accurate for the TA modes, likely due to the neglect of the piezoelectric long-range interaction, which is included in the EPW and ML calculations. 

All three methods used show that the largest scattering rate is due to LA phonons [Figs.~\ref{fig:dp} and \ref{fig:ML} (e)], exhibiting a weak energy dependence in the energy interval considered because it is proportional to the density of states. 
The scattering due to the polar optical (LO$_2$) mode is the second strongest scattering mechanism in the middle of the energy interval considered, with the characteristic energy dependence indicating the onset of phonon emission near 50 meV. The homopolar (ZO$_1$) phonon has a similar energy dependence compared to the polar optical phonon, but it is the weakest scattering mechanism overall. The scattering rates due to TA and TO$_2$ modes increase with energy, with the scattering rate due to  TO$_2$ mode becoming the second strongest scattering mechanism at higher energies. 

Finally, we compare the mobility calculated using the model and ML approaches with that obtained using the DFPT+Wannier approach, see Figs.~\ref{fig:dp} and \ref{fig:ML} (f).
It is clear that the ML approach is almost as accurate as the more rigorous DFPT+Wannier approach, while the model approach is somewhat less accurate. 
Furthermore, the ML approach is very easy to implement, unlike the model approach, which needs to be carefully formulated and parameters  calculated.

\textcolor{black}{The experimentally reported room-temperature mobilities for monolayer MoS$_2$ vary by an order of magnitude [between $\sim 15$ and $\sim 150$ cm$^2$/(Vs)]~\cite{Radisavljevic2013,zhihao16,Cui2015,Liu2015}. 
These discrepancies are commonly attributed to differences in the quality of samples, substrates and contacts. The room-temperature mobility obtained with our ML model and the EPW framework are 159 and 156 cm$^2$/(Vs), respectively, and are close to the maximal mobility value of  148 cm$^2$/(Vs) reported by Yu et al.~\cite{zhihao16}.}

\textcolor{black}{There have been several first-principles studies of the electron mobility of 2D MoS$_2$ in recent years~\cite{Kaasbjerg2012,Kaasbjerg2013,Li2013,Jin2014,Li2015,Gunst2016,Sohier2018,Guo2019,Gaddemane2021,Ponce2023_PRB,Ponce2023_PRL,Backman2024}. 
The reported first-principles values of mobility vary between 
$\sim 110$ and $\sim 410$ cm$^2$/(Vs).
This may stem from different exchange-correlation functionals leading to different energy differences between the K/K$^{\prime}$ and Q valleys (the Q valleys are closest in energy to the K/K$^{\prime}$ valleys), but also from different methods to calculate electron-phonon matrix elements (using supercells, linearly or Wannier interpolated DFPT).
The conceptual framework to accurately include the effect of long-range interactions in 2D materials was developed and applied to MoS$_2$ only in Refs.~\cite{Ponce2023_PRB,Ponce2023_PRL}. Ref.~\cite{Ponce2023_PRB} reported a room-temperature electron drift mobility of $132$ cm$^2$/(Vs), in good agreement with our results.}

\subsection{Thermoelectric transport properties of 2D MoS$_2$ at 300 K}\label{sec:TE}

\begin{figure}[h!]
\begin{centering}
\includegraphics[keepaspectratio, width=0.48\textwidth]{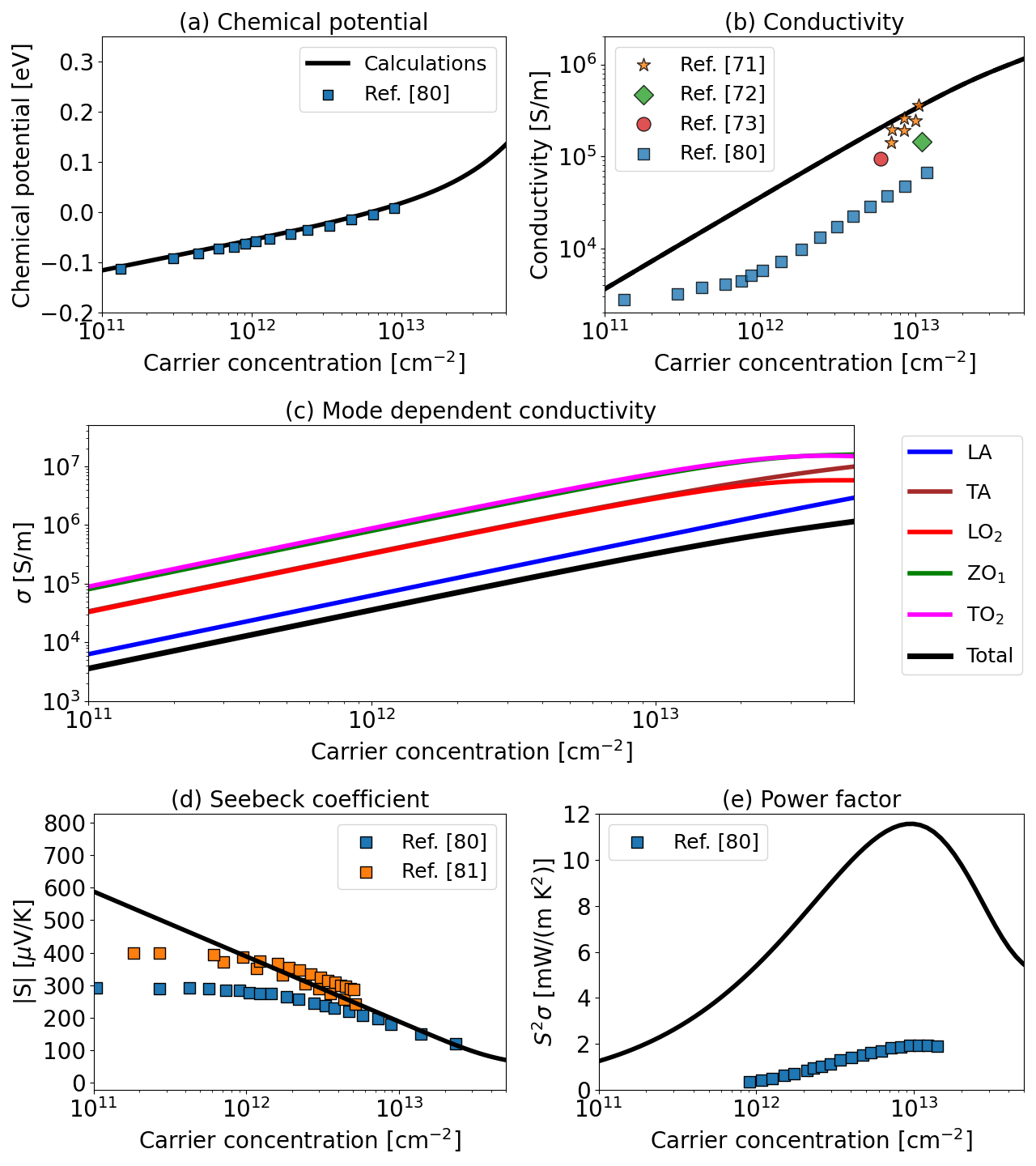}
\caption{\label{fig:TE_para_ML} Thermoelectric transport properties of monolayer MoS$_2$ at 300 K as a function of carrier concentration: (a) chemical potential, (b) electrical conductivity, (c) phonon mode-resolved conductivity, (d) Seebeck coefficient, and (e) power factor.  The solid lines correspond to our results using the machine-learning approach. The squares represent the experimental data taken from Ref. \cite{ng17,hippa17}.
\textcolor{black}{The other symbols represent estimates of conductivity taken from Refs.~\cite{zhihao16,Cui2015, Liu2015}.} 
}
\end{centering}
\end{figure}

In this section, we obtain the thermoelectric transport coefficients of 2D MoS$_2$ at room temperature and compare them with available experimental data~\textcolor{black}{\cite{ng17,hippa17,Cui2015, Liu2015,zhihao16}}.
Previous studies of thermoelectric transport in 2D MoS$_2$ either
did not properly account for the effect of long-range interactions on el-ph scattering or employed simpler scattering models than those developed in this work~\cite{Babaei2014,Jin2015,Zhao2018,Bilc2021,Wisesa2024,Kumari2024}. Furthermore, no comparisons were made with experimental data \textcolor{black}{on thermoelectric transport coefficients~\cite{ng17,hippa17}}.

Here we will present only the ML results, as the results of the model approach are very similar. 
Fig.~\ref{fig:TE_para_ML}~(a) illustrates our calculated chemical potential as a function of electron concentration and shows excellent agreement with the experimental data~\cite{ng17,hippa17}. 
In Fig.~\ref{fig:TE_para_ML}~(b), we compare the conductivity obtained using our ML approach and \textcolor{black}{measured or estimated experimentally \cite{ng17,hippa17,Cui2015, Liu2015,zhihao16}}.  We find the standard behavior seen in semiconductors, i.e.~there is an increase in conductivity due to the increase in carrier concentration. 
\textcolor{black}{Refs.~\cite{ng17,hippa17} reported the conductivity values in the units of S/m, obtained by taking the
effective monolayer thickness $h_z^{\text{eff}}=6.6$ \AA, which includes the bare S--S nuclear 
separation ($h_z = 3.13$ \AA) together with the finite 
spatial extent of the sulfur electron clouds on outer surfaces of the 
S--Mo--S monolayer. Therefore, we use the same value of the effective monolayer thickness as Refs.~\cite{ng17,hippa17} in our conductivity calculations. Moreover, we estimated the conductivity values from Refs.~\cite{Cui2015, Liu2015,zhihao16} using the measured mobility values, the quoted carrier concentration values and $h_z^{\text{eff}}=6.6$ \AA.
Similarly to the mobility values, there is almost an order of magnitude difference between the conductivity values obtained from experiments, reflecting the sample, substrate and contact quality. Our conductivity values are in very good agreement with experiments for the sample with the highest mobility~\cite{zhihao16}.}

\textcolor{black}{The discrepancy between our calculations with other experimental data can be attributed to extrinsic disorder effects,\footnote{\textcolor{black}{In principle, another potential source of discrepancy could be the use of the momentum relaxation time approximation instead of the numerical solution of the linearized BTE. The momentum relaxation time approximation is accurate when transport is dominated by quasi-elastic intravalley acoustic phonon scattering, as is the case here. Indeed, the momentum relaxation time approximation was shown to yield mobility values nearly identical to those obtained from the linearized BTE for phonon-limited transport in MoS$_2$~\cite{li15}.}} such as charged impurities and substrate-induced scattering. To assess this effect quantitatively, we model ionized impurity scattering using a screened Coulomb potential of an impurity as a perturbation (see Supplementary Material~\cite{supp}) and add it to phonon scattering via Matthiessen’s rule.
We consider impurity densities up to $2\times 10^{12}$ cm$^{-2}$, which is consistent with experimentally inferred values for monolayer MoS$_2$~\cite{zhihao16,Liu2024}.} \textcolor{black}{For impurity concentrations in this range, the calculated conductivities fall within the range of experimentally reported values (see Supplementary Material \cite{supp}).
}

In Fig.~\ref{fig:TE_para_ML} (c), we plot the conductivity resolved by each mirror-even phonon mode to determine the dominant contributions to the thermoelectric transport coefficients. We find that the LA phonon mode has the largest contribution throughout the whole range of carrier concentrations. At very high electron concentrations ( $\sim 5\times 10^{13}$ cm$^{-2}$), we find that the polar optical (LO$_2$) phonon has a comparable contribution to conductivity to that of LA phonon mode, while other modes contribute much less. These conclusions are in accordance with the plots of scattering rates shown in Figs.~\ref{fig:dp} and \ref{fig:ML} (e), which also show that LA and, to a much lesser extent, LO$_2$ modes are the largest contributors to transport in the range of Fermi energies corresponding to realistic carrier concentrations, see Fig.~\ref{fig:TE_para_ML} (a).

The absolute value of the Seebeck coefficient versus carrier concentration is illustrated in Fig.~\ref{fig:TE_para_ML} (d). We find the standard semiconductor behavior of Seebeck coefficient, i.e.~the absolute value of $S$ decreases with increasing carrier concentration.
For different experimental samples, we find different values of the Seebeck coefficient. 
For example, the highest measured Seebeck coefficient varies from $\sim$ 280 $\mu$V/K to $\sim$ 380 $\mu$V/K for the carrier concentrations of $\sim 10^{12}$ cm$^{-2}$ \cite{hippa17, ng17}. \textcolor{black}{Our predicted values of the Seebeck coefficient are within  
the experimental data for higher carrier concentrations 
($\sim 10^{12}$--$10^{13}$ cm$^{-2}$), as seen in Fig.~\ref{fig:TE_para_ML} (d). 
However, at lower carrier concentrations ($<10^{12}$ cm$^{-2}$), our 
results overestimate the measured Seebeck coefficient. A likely reason for this is that our calculations consider only the lowest conduction 
band contribution to transport, whereas at such low carrier concentrations, 
thermally activated holes from the highest valence band may also contribute to 
transport, effectively reducing the total Seebeck coefficient. Ionized impurity scattering has almost negligible effect on the calculated Seebeck coefficient values.} 

The power factor as a function of the carrier concentration is shown in Fig.~\ref{fig:TE_para_ML} (e).
Ng {\it et al.}~\cite{ng17} measured the highest power factor of $S^2\sigma |_{exp} = 2.0$ $\times 10^{-3}$ W/(mK$^2$) with the optimal carrier concentration of $n\sim$ 1.3$\times 10^{13}$ cm$^{-2}$, while Hippalgaonkar  {\it et al.}~\cite{hippa17} measured the highest power factor to be $S^2\sigma |_{exp} = 3.3$ $\times 10^{-3}$ W/(mK$^2$). 
Our calculations predict the maximal power factor of 
$11.6$~mW\,K$^{-2}$m$^{-1}$ (for the effective thickness of $h_z^{\text{eff}}=6.6$~\AA) with the optimal carrier concentration of 
$n = 10^{13}$ cm$^{-2}$. 
When ionized impurity 
scattering is included (see Supplementary Material~\cite{supp} ), with impurity densities up to $2\times 10^{12}$ cm$^{-2}$, 
the maximal power factor is achieved for carrier concentrations in the range of
$(1$--$2)\times10^{13}$ cm$^{-2}$, with peak values ranging from 
$2$ to $11.6$ mW\,K$^{-2}$m$^{-1}$. At high impurity concentrations, the calculated power factor shows good 
agreement with experimentally measured values, confirming that impurity 
scattering is likely the dominant mechanism limiting thermoelectric 
performance in these samples (see Supplementary Material~\cite{supp}).

Importantly, the \textcolor{black}{predicted phonon-limited} room temperature power factor of single layer MoS$_2$ is larger than those of the commercial thermoelectric materials bismuth telluride~\cite{Witting2019,Cha2019} and lead telluride~\cite{Pei2012,Vineis2008,Jood2020} \textcolor{black}{by a factor of $\sim 2-4$}. The large power factor can be attributed to the large effective mass of monolayer MoS$_2$ and its 2D density of states. Heterostructures of MoS$_2$ have been shown to enhance the power factor~\cite{ng17,hippa17} and reduce the lattice thermal conductivity \cite{Kim2021}, 
so they are promising candidates for next generation thermoelectric generators.

\section{Summary and conclusions}

 We have developed \textcolor{black}{computationally inexpensive} model and machine-learning approaches combined with first-principles calculations that provide an accurate description of electron-phonon interaction in semiconductors. We have applied these methods to  calculate thermoelectric transport properties of two-dimensional MoS$_2$. The model approach includes the Fr\"ohlich model to describe polar optical modes, while the deformation potential models  describe electron-phonon scattering processes due to other modes. The new aspect of our model approach is the generalization of the acoustic deformation potential approach for arbitrary crystal symmetries and arbitrary locations of band extrema. All the parameters for the model approach can be computed using state-of-the-art density functional perturbation theory (DFPT).
 In the machine-learning approach, we use regression to obtain electron-phonon matrix elements on dense wave vector grids in the regions of the Brillouin zone relevant for transport starting from relatively small wave vector grids obtained using DFPT.

 We have shown that both the model and machine-learning approaches show very good agreement with the DFPT+Wannier approach for dense wave vector grids.
The thermoelectric transport coefficients derived from both models are also in good agreement with available experimental data and accurately capture the high thermoelectric power factor of 2D MoS$_2$. We find that the dominant scattering mechanism 
is due to longitudinal acoustic phonon modes.

We highlight that both the model and machine-learning approaches can predict accurate thermoelectric transport coefficients for a variety of materials at a much reduced computational time compared to the standard DFPT+Wannier approach. \textcolor{black}{In this work, the electron-phonon matrix elements in both model and machine learning approaches were obtained from the DFPT+Wannier calculations primarily because such a benchmark was needed for the validation of both approaches. However, neither method intrinsically requires the Wannier interpolation: the relevant parameters can be extracted directly from DFPT calculations using a small number of phonon wave vectors, which can drastically reduce the computational cost.}
The ML approach \textcolor{black}{can be two orders of magnitude  faster} than the DFPT+Wannier approach and has comparable accuracy. The ML approach is also straightforward to implement.
The model approach is less accurate than the DFPT+Wannier and ML approaches (but still fairly accurate) \textcolor{black}{and can be three orders of magnitude faster} than the DFPT+Wannier approach. \textcolor{black}{However, the model approach} may require considerable effort to be formulated for each material of interest. Therefore, the ML approach may be the most advantageous method among the three for screening different material systems. 

\section*{Acknowledgement}
We thank Stephen Fahy, Samuel Ponc\'e and Yunhao Mo for helpful discussions.
This project received funding from the European Union’s Horizon 2020 research and innovation programme under the Marie Skłodowska-Curie grant agreement number 713567. This work was partly supported by Science Foundation Ireland under grant numbers 15/IA/3160 and 13/RC/2077. The later grant was co-funded under the European Regional Development Fund. We acknowledge the Irish Centre for High-End Computing (ICHEC) for the provision of computational facilities. R. D. acknowledges CSC-IT Center for Science in Finland for supporting this work with high-performance computing resources.

\appendix

\section{Intravalley non-polar optical and intervalley deformation potentials\label{ap: in_ac_dp}}

To obtain the independent DP components, we adopt the method described by Murphy {\it et al.} \cite{Murphy2018} and D'Souza {\it et al.} \cite{DSouza2020}. In this method, we relate the el-ph matrix elements obtained from DFPT and the Wannier interpolation with deformation potentials along high-symmetry lines using Eq.~\eqref{eq:ac} and  Table \ref{tab:LaTa-cont}. 

In DFPT, the el-ph matrix element  from a state $\bm{k}$ and band index $n$ to a state $\bm{k+q}$ and band index $n'$ is given as \cite{giustino17}
\begin{eqnarray}\label{eq:me_dfpt}
g_{n{\bm k},\lambda{\bm q}}^{n'\bm {k+q}} = \sqrt{\frac{\hbar}{2\omega_{\lambda\bm{q}}}}\sum_{b,i}\sqrt{\frac{1}{m_b}}e^{\bm{q}\lambda}_{b,i} \nonumber \\
\times \bra{u_{n'\bm{k+q}}} \partial_{b,i,\bm{q}}v^{\rm{ks}} \ket{u_{n\bm{k}}}, 
\end{eqnarray}
where $u_{n\bm k}$ is the lattice periodic part of the wave function given as $\frac{1}{\sqrt{N_l}}u_{n\bm k}e^{i\bm{k\cdot r}}$, where $N_l$ is the number of primitive cells. $e^{\bm{q}\lambda}_{b,i}$ is the $i^{\rm{th}}$ Cartesian component of the phonon eigenvector  associated with atom $b$ of mass $m_b$. $\bm q$ and $\lambda$ are the wave vector and branch of the phonon with frequency $\omega_{\lambda\bm{q}}$ that is involved in the scattering event. $\partial_{b,i,\bm{q}}v^{\rm{ks}}$ is the lattice periodic part of the first-order expansion of the perturbed Kohn-Sham potential \cite{giustino17}. 
  
\begin{figure}[h!]
\begin{centering}
\includegraphics[keepaspectratio, width=0.45\textwidth]{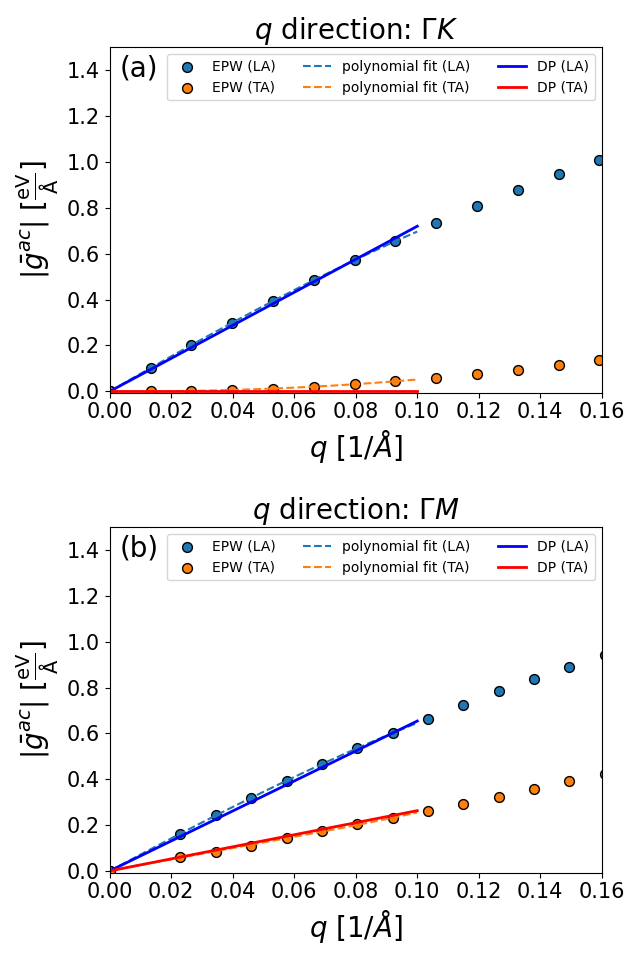}
\caption{\label{fig_ap:dp_2d} Scaled electron-phonon matrix elements due to acoustic modes between the electronic states at K and K+${\bm q}$ as a function of the phonon wave vector ${\bm q}$ whose directions are indicated at the top of each panel. Orange and blue dots represent the matrix elements calculated using density functional perturbation theory and the Wannier interpolation for transverse and longitudinal acoustic modes, respectively. Dashed lines are the third-order polynomial fits to those matrix elements. Solid lines are obtained using the computed values of deformation potentials (DPs) given in Table \ref{tab:LaTa-cont} for transverse and longitudinal acoustic modes in the long-wavelength limit.}

\end{centering}
\end{figure}

The el-ph matrix elements are calculated using DFPT and the Wannier interpolation between the electronic states at K and K$+{\bm q}$ using Eq.~(\ref{eq:me_dfpt}) as $|{\bm q}| \rightarrow 0$ along two directions, ${\bm q} \parallel$ to $\Gamma$-K and ${\bm q} \parallel$ to $\Gamma$-M. The DFPT electron-phonon matrix elements divided by the phonon amplitude $\sqrt{\hbar/2m \omega_{\lambda{\bm q}}}$ are then plotted and fitted to a third-order polynomial, see Fig.~\ref{fig_ap:dp_2d}. We fit the linear coefficient terms to the expressions given by Eq.~\eqref{eq:ac} and Table \ref{tab:LaTa-cont} using a linear regression technique and obtain the linearly independent deformation potentials for the K valley. The calculated independent components are shown in Table \ref{tab:dp}. It is clear from Fig. \ref{fig_ap:dp_2d} that in the long wavelength limit (${\bm q}\rightarrow 0$), analytical expressions accurately describe the first principle matrix elements. We note that when ${\bm q}$ is parallel to $\Gamma$-K, the deformation potential due to the TA phonon mode vanishes.
However, the non-linear coefficients of the DFPT+Wannier matrix elements do not vanish, and they are the consequence of long-range piezoelectric interaction.

\begin{thebibliography}{10}
\bibitem{mos2phonon}
The experimental points correspond to bulk MoS$_2$, but only for the modes that are expected to appear in the monolayer.
\end{thebibliography}
\end{document}